**Toward a Cognitive Data Model: Exploring a Mind-Inspired Approach to Database Design**

Dhammika Pieris

*Department of Information Systems Engineering and Informatics*

*Faculty of Computing*

*University of Sri Jayawardenepura, Sri Lanka*
dhammikapieris@sjp.ac.lk

1. **Abstract**

The Cognitive Data Model (CDM) is proposed. A novel approach to database design, inspired by the belief that the human brain operates with a logical data model independent of its anatomical structure. The study aims to identify and replicate this CDM to enhance database design, resolving limitations of the existing models in handling complex relationships, scalability, and adaptability.

The methodology involves empirical observation, cognitive experiments, iterative modelling and critical thinking. Findings suggest that the information processing in the brain occurs sequentially with forming associations on atomic static data in a pairwise, time-dependent manner. This insight led to the development of a meta-framework for the brain's information processing to design CDM based on it.

The CDM offers improved data modelling and database design, efficient querying, and potential applications in AI, machine learning, and big data analytics. Future research will focus on formalizing the model, implementing it in 3D environments, and developing a query language.

## 2. Introduction

Data models play a foundational role in database design and management. The relational data model, introduced by Codd (1970), provides a structured approach to modelling, storing, and querying data. It ensures that each attribute holds indivisible (atomic) values, facilitating efficient data organisation and retrieval. However, despite its widespread adoption, the relational model has limitations in handling complex relationships, large-scale data structures, diverse datasets (Mukala, 2025), and massive amounts of unstructured data (Miryala, 2024). To address these challenges, alternative models—such as document, key-value, wide-column, and graph data models—have emerged, collectively known as NoSQL databases.

While these models offer certain advantages, the relational model remains the de facto standard in many application domains — e.g., financial and banking systems, enterprise resource planning (ERP), healthcare records, government and public sector databases, supply chain and inventory management, airline and railway reservation systems, and telecommunications and billing systems (Atlan, 2024). These domains demand strict data integrity, consistency, concurrency control, transaction reliability, and robust security and access control.

This research aims to find a data model that preserves the strengths of the relational model—such as data integrity, consistency, and security—while addressing its limitations in handling complex relationships and large-scale data structures.

Data models are human creations, often based on individual experiences and perceptions about the real world. People comprehend and interact with the world through their minds, which suggests that there may be a logical way of organising information in the brain independent of its anatomical structure.

Human memory and cognitive processing, however, rely on the brain's anatomical structure. For instance, the hippocampus, prefrontal cortex, and thalamus are three key regions in the brain that support memory and cognitive processing. The hippocampus converts short-term memories into long-term storage and aids spatial memory(Wang, 2025). The prefrontal cortex manages working memory, decision-making, and cognitive control over stored information(Funahashi, 2017; Levy, 2023). While the thalamus acts as a relay centre, transmitting sensory and cognitive data between brain regions (Mitchell *et al.*, 2014). These structures function together, showing that memory, reasoning, and data communication are physically embedded in the brain's organisation, reinforcing the dependence of human cognition on its anatomical structure.

Building on this idea, the current research hypothesizes that such memory and cognitive processing, while dependent on the brain's structure, lead to the existence of a logical data model that is independent of this organisation. If such a cognitive data model could be identified, it might provide a new foundation for designing computerised databases. The study aims to reconstruct this model within computational systems, developing a versatile, efficient, and scalable approach to database design.

Several data models have been proposed as representations of the human brain's data model – e.g., Nural Networks (Andina *et al.*, 2007; Chen *et al.*, 2008) . Most of these models incorporate both static data (stored knowledge representations)(Savarimuthu *et al.*, 2024; Sridhar *et al.*, 2023) and dynamic data (which captures events and cognitive processing over time)(Zacks *et al.*, 2007). No existing brain-inspired data model appears to focus exclusively on granular-level atomic static data—such as Car, Country, Chair, or Furniture—as standalone entities connected by simple relationships. Instead, existing models incorporate attributes, complex relationships, and processing mechanisms, and they are invariably tied to the brain's anatomical structure.

This research hypothesizes that such a purely static and granular-level atomic data model exists in the human brain as a logical data model independent of the brain's anatomical structure. The study aims to mimic this model and implement it in a computational system to be used for database development. Given the structured nature of this proposed model, it is believed to be well-suited for transaction-oriented databases, such as financial, banking, and airline reservation systems, which rely on clear, consistent, efficient, and structured organisation of data—characteristics that are inherent in relational databases.

## 3. Methodology

The research commences with a theoretical framework hypothesizing that the human brain uses a logical data model independent of its anatomy. To invent this model, the study employs a combination of empirical observation, critical thinking, personal beliefs, life experiences, interviews, and iterative modelling. Cognitively engaged with a diverse group of individuals to understand their thought processes and perceptions of real-world objects and events. Participants were selected through convenience sampling, and no rigid criteria were imposed to allow diverse perspectives.

One of the experiments conducted aimed to determine whether a participant could identify two things simultaneously or with a time gap. For example, a pen clip and an eraser were placed on a table, and the participant was asked to identify them. Another approach involved asking whether a person could recognize two things presented at the same time. In both cases, the respondent's ability to identify two or more objects was discussed and analyzed through direct engagement. The findings indicated that all identified objects were recognized sequentially rather than simultaneously. By analogy, it was inferred that any information-processing event in the human brain occurs sequentially, one after the other. Another idea explored was if multiple pieces of information are to be associated in the human brain, whether they are associated with each other in the brain at the same time.

For experimenting with this idea, another question was raised: Could three people (e.g., friends) meet at the same place, on the same day, and at the same time? Initially, respondents sought clarification on what was meant by "meet." After discussion, the respondents agreed that meeting was the moment of first eye contact. The results revealed that only two people could make eye contact at the exact same moment. Then, one of them could make eye contact with the third person, followed by the other.

By analogy, it was inferred that only two things can be associated at a time in the human brain. Once an association is formed, only one associated element can connect with a third, and this process continues sequentially.

Based on these findings, it was hypothesized that conceptual models (e.g., the Entity Relationship (ER) model(Chen, 1976)) and logical database models (e.g., the relational database schema) should be designed analogously to how the human brain processes information. That is, the brain identifies things sequentially and associates them in pairs, also in a sequential manner.

Rather than modifying existing data models (e.g., the relational database schema), developing a new data model aligned with these observed cognitive processes might be more promising. This realization led to an exploration of a data model that mirrors the brain's information-processing behaviour, marking the beginning of the cognitive data model.

During the development of the model, nearly two hundred real-world database scenarios—sourced from textbooks, research papers, and the Internet—were iteratively mapped to the emerging model to hypothesize, refine, and evaluate its structure. The evaluation process followed a trial-and-error approach, incorporating feedback from cognitive experiments to ensure alignment with observed cognitive processes.

Years of continuous exploration discovered that while the cognitive data model could be represented graphically as a diagram on paper, it alone was insufficient to convey its principles to others effectively. A gap existed between the findings on how the brain processes information and the graphical model designed to mimic this process. While the diagram could be interpreted, explaining it to others proved challenging. To bridge this gap, it was realized that a meta-framework was needed—one structured based on the observed principles of the brain's information processing. Rather than constructing the cognitive data model directly from experimental findings, it had to be developed within this framework to ensure clarity, coherence, and applicability. This realization marked a pivotal shift in the research, leading to the formalization of a foundational framework for the CDM.

## 4. The Foundational Framework for the CDM

After nearly a decade of rigorous research and exploration, a meta-level framework—known as the Cognitive Data Model Framework (CDMF)—has been developed to conceptualize the anticipated Cognitive Data Model (CDM). Outlined below, this framework represents a foundational breakthrough in understanding how the human mind logically structures and associates information at a granular level. The CDMF consists of a five-step procedure as detailed below.

(1) A thing in the human mind refers to anything that carries significant real-world meaning as perceived by a person and can be expressed in words—for example, car, green colour, red big onion, or Olympics 2024.

(2)

    (a) A thing can be associated with another thing to enhance knowledge about it. For example, 'Chair' can be associated with 'Furniture' to indicate that a Chair is Furniture. Similarly, 'Mango' can be associated with 'Fruit' to indicate that a 'Mango' is a 'Fruit.'

    (b) In an association between two things, one acts as the *member* and the other as the *owner*. The *member* determines the *owner* and is called the *determinant* or *determiner*, while the *owner* qualifies the *member* and is called the *qualifier*. For instance, in the association between 'Chair' and 'Furniture', 'Chair' is the member (or determinant), and 'Furniture' is the owner (or qualifier). The 'Chair' determines 'Furniture', and 'Furniture' qualifies 'Chair'.

    (c) Associations follow a *member-to-owner* structure, meaning that a participating thing can play only one role—either *member* or *owner*—but not both in the same association. Thus, associations are unidirectional.

(3)

    (a) Each thing in the mind has a time attribute that continuously updates as long as it exists.

- (b) When a thing is associated with another, the exact date and time of the association are recorded, along with a reference to the associated thing. The recorded time reflects the exact moment the association took place.

- (c) However, the thing's time attribute continues to progress independently.

(4)

- (a) An association between two things is an *event* called an association event. An association event always occurs between exactly two things.

- (b) A thing already associated with another thing does not associate with another third thing at the same time.

- (c) No two associations occur at the same time, even between different pairs of things

(5) A thing can be associated with multiple things but only through separate association events, each occurring between two things at distinct times sequentially, one after another, with a time gap between any two association events.

The CDMF's five steps can be called Step (1), Step (2)-(a), …, and Step (5). The insights gained from the CDM Framework (CDMF)—the meta-schema—serve as the foundation for developing the Cognitive Data Model (CDM), which will be explained in the following sections.

## 5. Building the CDM

According to Step (1) of the CDMF, things exist in the human mind and can be identified using words or phrases, such as 'Chair,' 'Furniture,' and 'Mount Everest.' Based on this, it can be assumed that 'Chair' and 'Furniture' exist in the mind. Following Step (2)-(a), it is assumed that 'Chair' associates with 'Furniture' to enhance knowledge. Step (2)-(b) classifies 'Chair' as the member (or determinant) and 'Furniture' as the owner (or qualifier). According to Step (2)-(c), this is a unidirectional member-to-owner association.

Per Step (3)-(a), both 'Chair' and 'Furniture' have a time attribute that continuously updates as long as they exist. Step (3)-(b) defines $t_0$ as the actual time when 'Chair' and 'Furniture' associate and this time is recorded.

This association is graphically represented by an arrow from 'Chair' (member) to 'Furniture' (owner) in Figure 1.

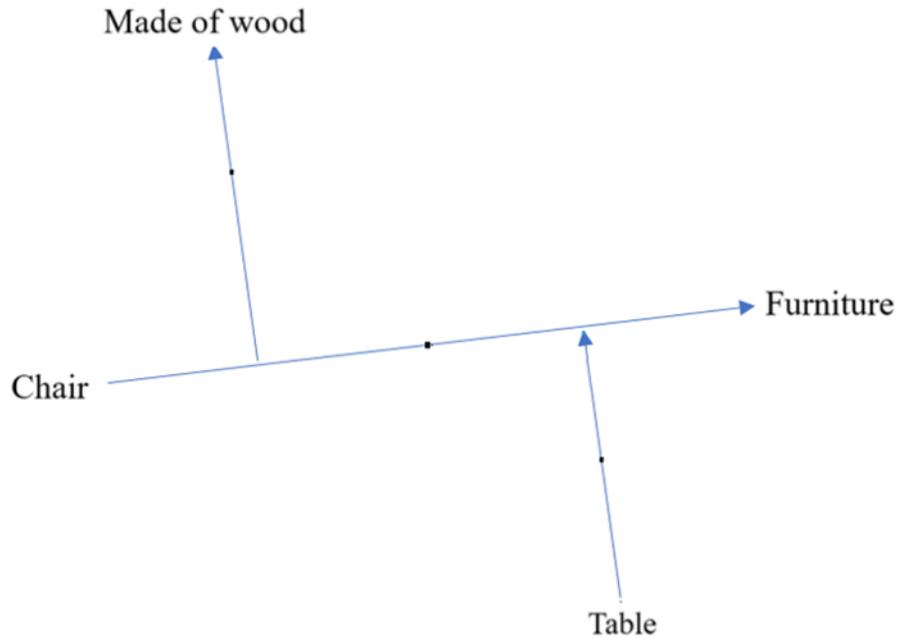

Figure 1: A mini–Cognitive Data Model (CDM)

The things 'Chair' and 'Furniture', which are at the time $t_0$, when the association is created, are shown in Figure 1 by the texts 'Chair' and 'Furniture' themselves that appear on either side of the arrow. Notice that the time is not shown on the diagram.

From this point onward, we denote a thing at a time $t_0$ using the notation 'Chair' $t_0$ and 'Furniture' $t_0$.

Following Step (3)-(c), the time attribute of both 'Chair' and 'Furniture' continues to progress independently after the association. Even though 'Chair' and 'Furniture' have already been associated, Step (5) allows them to form new associations but with different things and at different times. To reflect this, we introduce two new different time variables: $t_1$ for 'Chair' and $t_2$ for 'Furniture.' Initially, as long as neither 'Chair' $t_1$ nor 'Furniture' $t_2$ associates with anything else, $t_1$ remains equal to $t_2$.

Since 'Chair' $t_0$ and 'Furniture' $t_0$ have already been associated, Steps (4)-(a) and (4)-(b) state that neither can associate with a third thing. To depict the present state of 'Chair' $t_1$ and 'Furniture' $t_2$

, the arrow from 'Chair' to 'Furniture' is divided into two parts by a dot in the middle. The tail represents 'Chair' $t_1$, while the head represents 'Furniture' $t_2$.

Beyond 'Chair' and 'Furniture,' many other things exist in the mind. Returning to Step (1), we introduce 'Made of wood.' Since 'Chair' $t_0$ no longer participates in any new associations, Step (5) allows the 'Chair' $t_1$ to associate with 'Made of wood.' Given that $t_1$ is the current time, we assume 'Made of wood' also exists at $t_1$, forming the association 'Chair' $t_1 \rightarrow$ 'Made of wood' $t_1$.

Graphically, 'Chair' $t_1$ is already represented by the tail of the first arrow. Thus, we add a second arrow from this tail to 'Made of wood' $t_1$. This second association also follows the member-owner structure, with 'Chair' $t_1$ as the member and 'Made of wood' $t_1$ as the owner.

Per Step (3)-(c), after this association forms, both things update to a new, later time value. We denote these as $t_3$ for 'Chair' and $t_4$ for 'Made of wood.' In Figure 1, 'Chair' $t_3$ is represented by the tail of the second arrow, while 'Made of wood' $t_4$ is represented by the arrowhead. Both are now available for new associations. However, per Step (4)-(c), associations must occur sequentially, leading to distinct time values $t_3$ and $t_4$. As long as neither 'Chair' $t_3$ nor 'Made of wood' $t_4$ forms another association, we maintain $t_3 = t_4$.

Similarly, we introduce 'Table' and associate it with 'Furniture' $t_2$, denoting it as 'Table' $t_2$. In this third association, 'Table' $t_2$ is the member, and 'Furniture' $t_2$ is the owner. A third arrow is added to Figure 1, pointing from 'Table' $t_2$ to 'Furniture' $t_2$. The arrow's tail represents 'Table' $t_6$, while the head represents 'Furniture' $t_5$, both of which are now available for further associations.

When the number of things and associations is relatively small, the model can be depicted in two dimensions on paper using a pen or pencil (Figure 1). Thus, Figure 1 illustrates a smaller CDM that can be represented in two-dimensional space.

The CDM in Figure 1 functions as a data model, meaning the things, such as 'Chair' and 'Furniture,' can also be referred to as data items. The model in Figure 1, though small and seemingly simple, adheres to all five steps of the CDMF framework. It could not have been created without following each step.

The real-world scenario represented by the mini-CDM (Figure 1) can be summarized as: A 'Chair' and a 'Table' are types of 'Furniture', and a 'Chair' is 'Made of wood'.

Figure 2 below contains more things and associations and shows an extended version of the CDM exposed in Figure 1.

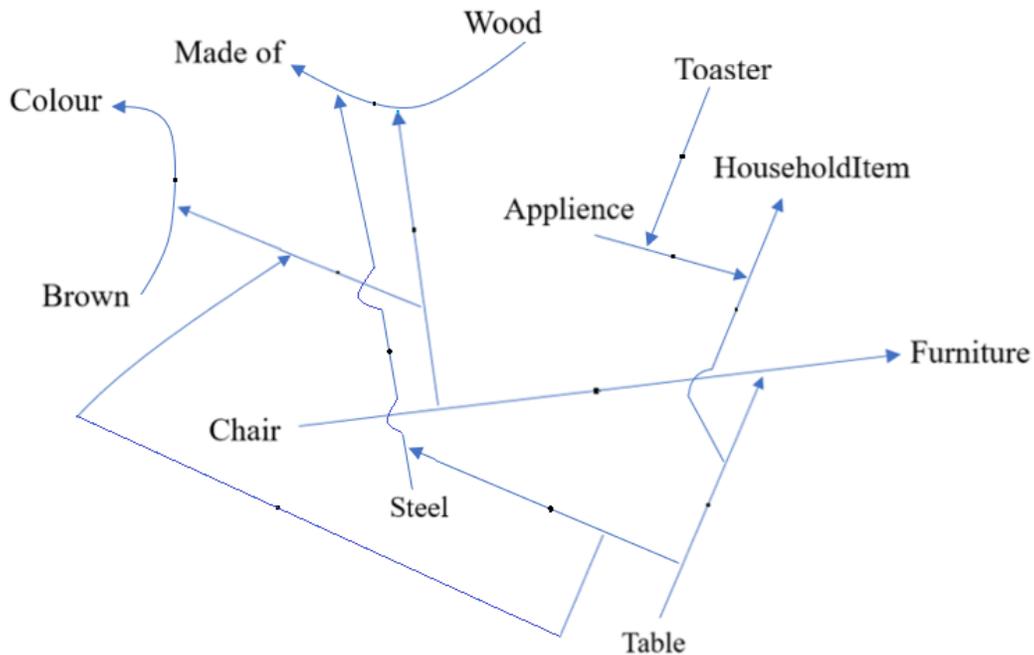

Figure 2: A large CDM - an extended version of the model in Figure 1

The real-world scenario represented by the model in Figure 2 can be described as follows: A 'Chair' and a 'Table' are types of 'Furniture', while a 'Toaster' is a type of 'Appliance'. Both 'Appliance' and 'Furniture' fall under the category of 'HouseHoldItem'. The 'Chair' is 'Made of' 'Wood' and is 'Brown' in 'Colour', while the 'Table' is 'Made of' 'Steel' and is also 'Brown' in 'Colour'.

However, the model does not represent the toaster's material and colour. This could be because they may be inapplicable, unknown, or simply missing. Nevertheless, the situation can be managed without using negation or null values – an advantage not available in the relational data model.

If needed, these attributes could be incorporated by extending the model further. However, doing so would be challenging, as the current model is already congested within the constraints of a two-dimensional space. Given the increasing number of entities and associations, a three-dimensional space may be required for a more effective representation. Further research is needed to explore computational methods—such as 3D graphics—for implementing a three-dimensional model. Additionally, future studies should focus on converting this computational model into a structured data model that can serve as the foundation for database design.

## 6. Discussion

The development of a Cognitive Data Model could significantly transform database technology. By emulating the hypothesized data representation mechanism of the human brain, this model is expected to enhance data integrity and consistency. Additionally, it aims to simplify complex relationships, reduce system complexity, improve query performance, and drive innovation in AI, machine learning, and big data analytics.

The study identified the most granular atomic-level logical data model used by the human brain, which can be described independently of its anatomical structure. Although previous studies, such as the Conceptual Micro-Object Model (CMoM)(Chua *et al.*, 2002), have explored attribute-centric data modelling, these models still acknowledge some implicit segmentation of entity types, similar to ER models.

In contrast, this research proposes a model where attributes and their values are not inherently distinct in structure. An attribute value in one association can function as an attribute in another, and vice versa, depending on the context. If we view tables, rows, columns, attributes, and attribute values as hierarchical levels of data organization, this research delves into the lowest level—attribute values. It introduces a data model where attribute values themselves serve as fundamental data items, forming the building blocks of the model.

Figure 1 illustrates, while Figure 2 more clearly demonstrates, that each value (referred to as a *thing*) connects to only one other thing at a time. If it needs to connect to multiple things, it does so at different points in time. Due to these time-varying association features and the adoption of a graphical, human-accessible model (as shown in Figures 1 and 2), these values (or data items) remain traceable.

Any data item that serves as either a member or an owner of another can be modelled by the CDM. It is believed that every data item is either a member or an owner of another. Therefore, the CDM can represent anything in the world that can be expressed through language.

Regardless of the model's size, a unique path exists between any two data items, as evident in Figures 1 and 2. Furthermore, all data items are connected in pairwise, unidirectional, member-to-owner relationships, making them easily searchable and individually updatable without affecting other items or connections. New items can be freely added, and existing items can be removed without disruption.

This flexibility opens up limitless opportunities for experimentation and innovation. Consequently, several promising future research directions can be explored as follows:

   I. Implementing the CDM in 3D graphics and transforming it into a database system.

  II. Demonstrating the CDM's advantages over existing data models in efficiency, scalability, and flexibility.

 III. Formalizing the CDM using alphanumeric representations instead of real-world names (e.g., Chair, Furniture, Toaster) as shown in Figure 2.

  IV. Developing a procedure to transform the CDM into a relational data model.

   V. Designing a query language for the CDM to support data creation, control, and manipulation.

  VI. Developing a conceptual model, similar to Chen's ER model, to represent real-world scenarios and facilitate their transformation into the CDM, and

 VII. Investigating the Potential Applications of CDM in AI, Machine Learning, and Big Data Analytics.

## 7. Conclusion

The primary objectives of this study were to identify the hypothesized granular atomic logical data model that the human brain uses to process and represent information. Subsequently, the study aimed to design a computational equivalent and validate its effectiveness in database design and application scenarios.

This research bridges the gap between human cognition and logical data modelling for database development, offering a transformative approach to database design. The proposed Cognitive Data Model has the potential to revolutionize how databases are conceptualized, developed, and applied across industries.

## 8. Acknowledgement

The author acknowledges the use of generative AI, ChatGPT, as a writing assistant in the preparation of this paper. ChatGPT was employed to refine paragraphs, improve grammar, enhance clarity, summarize content, ensure logical flow, and assist in organizing thoughts. However, all intellectual contributions, research findings, and conclusions presented in this paper are solely the author's own.

This research was conducted relying solely on the author's personal funds. The author has applied for external funding but has not received it.

## 9. References

Andina, D.*, et al.* (2007). Neural networks historical review. 39-65.

Atlan. (2024). Relational vs NoSQL databases: Which is right for you in 2024? Retrieved from https://atlan.com/relational-database-vs-nosql/

Chen, J.*, et al.* (2008). *Data-brain modeling based on brain informatics methodology.* Paper presented at the 2008 IEEE/WIC/ACM International Conference on Web Intelligence and Intelligent Agent Technology.

Chen, P. P. S. (1976). The entity-relationship model: toward a unified view of data. *ACM Trans. Database Syst, 1*(1), 9-36. doi:10.1145/320434.320440

Chua, C. E.*, et al.* (2002). On conceptual micro-object modeling. *Journal of Database Management (JDM), 13*(3), 1-16.

Codd, E. F. (1970). A relational model of data for large shared data banks. *Commun. ACM, 13*(6), 377-387. doi:10.1145/362384.362685

Funahashi, S. (2017). Working memory in the prefrontal cortex. *7*(5), 49.

Levy, R. (2023). The prefrontal cortex: from monkey to man. *Brain, 147*(3), 794-815.

Miryala, N. K. (2024). Emerging Trends and Challenges in Modern Database Technologies: A Comprehensive Analysis.

Mitchell, A. S.*, et al.* (2014). Advances in understanding mechanisms of thalamic relays in cognition and behavior. *34*(46), 15340-15346.

Mukala, P. (2025). Bridging the Gap: Pre-Hadoop Relational Systems and the Evolution of Emerging Data Technologies.

Savarimuthu, A.*, et al.* (2024). Receive, retain and retrieve: psychological and neurobiological perspectives on memory retrieval. *58*(1), 303-318.

Sridhar, S.*, et al.* (2023). Cognitive neuroscience perspective on memory: overview and summary. *17*. doi:10.3389/fnhum.2023.1217093

This is about how brain store memories

Wang, H. (2025). Behavioral decision-making of mobile robots simulating the memory consolidation mechanism of human brain. *33*(2), 135-149. doi:10.1177/10597123241302508

Zacks, J. M*., et al.* (2007). Event perception: a mind-brain perspective. *133 2*, 273-293.